\documentclass[12pt]{article}

\renewcommand{\d}{{\rm d}}

\newcommand{\M}{{\cal M}}
\newcommand{\R}{{\cal R}}
\newcommand{\D}{{\cal D}}

\begin{document}

\title{GL(4,R) representation of the gravity action on the piecewise flat spacetime
}

\author{V.M. Khatsymovsky \\
 {\em Budker Institute of Nuclear Physics} \\ {\em of Siberian Branch Russian Academy of Sciences} \\ {\em
 Novosibirsk,
 630090,
 Russia}
\\ {\em E-mail address: khatsym@gmail.com}}
\date{}
\maketitle
\begin{abstract}
The gravity action on the piecewise flat Riemannian manifold is formulated using the discrete set of the nondegenerate 4$\times$4 matrices on the 3-simplices as some connection type variables. These variables are the discrete counterpart of the affine (Christoffel) connection used as independent variables in the Palatini form of the Einstein gravity action. Excluding these with the help of the equations of motion we get the original discrete gravity action on the piecewise flat spacetime (Regge action). The discrete version of the diffeomorphisms and path integral are briefly discussed.

\end{abstract}


MSC classes: 83C27; 53C05

keywords: Einstein theory of gravity; minisuperspace model; piecewise flat spacetime; Regge calculus; affine connection; Palatini action; discrete connection

There is a well-known property of the gravity Einstein action that it can be rewritten in terms of the metric and affine (Christoffel) connection variables,
\begin{eqnarray}\label{Palatini0}                                            
S = \int R^\nu_{\lambda \nu \mu} g^{\lambda \mu} \sqrt {g} \d^4 x = \int [ - \Gamma^\nu_{\lambda \mu} \partial_\nu (\sqrt{g} g^{\lambda \mu}) \nonumber \\ + \Gamma^\nu_{\lambda \nu} \partial_\mu (\sqrt{g} g^{\lambda \mu}) + \sqrt{g} g^{\lambda \mu} ( \Gamma^\nu_{\lambda \mu} \Gamma^\rho_{\nu \rho} - \Gamma^\nu_{\lambda \rho} \Gamma^\rho_{\nu \mu} ) ]  \d^4 x,
\end{eqnarray}

\noindent so that if the connection variables are treated as independent ones and excluded by using the equations of motion the original Einstein action in terms of the metric is restored (Palatini, \cite{Pal}). Here, $g \equiv \det \| g_{\lambda \mu} \|$ and we assume the Euclidean metric signature for definiteness. $g_{\lambda \mu}$ and $\Gamma^\lambda_{\mu \nu}$ are independent variables. An additional requirement for that the equations of motion would give for $\Gamma^\lambda_{\mu \nu}$ the unique metric-compatible connection for $g_{\lambda \mu}$ is the requirement that the connection be torsion-free, $\Gamma^\lambda_{\mu \nu} = \Gamma^\lambda_{\nu \mu}$.

Let us consider this on the piecewise flat spacetime geometry. This class of geometries is sufficiently large to approximate (in a certain topology) any metric with any accuracy \cite{CMS}. Such a spacetime can be taken as the simplicial complex or the set of the flat four-dimensional tetrahedra or 4-simplices $\sigma^4$. Generally, in the $d$-dimensional case, the $d - 2$-dimensional subsimplices $\sigma^{(d - 2)}$ support the curvature and are the sets of points of some conical singularities. The conical singularity is characterized by the defect angle $\alpha_{\sigma^{(d - 2)}}$ which is the difference of $2 \pi$ and the sum of the dihedral angles meeting at the given $(d - 2)$-simplex or triangle $\sigma^2$ in the physical case $d = 4$. Let $A_{\sigma^2}$ be the area of this triangle. Then the usual metric gravity action on such a spacetime reads
\begin{equation}                                                            
\frac{1}{2} \int R \sqrt{g} {\rm d}^4 x = \sum_{\sigma^2}A_{\sigma^2} \alpha_{\sigma^2}
\end{equation}

\noindent (Regge action, \cite{Reg}).

Using the discrete or lattice methods can help in quantising the formally nonrenormalizable general relativity (GR) (\cite{RW}, \cite{Ham}). So we aim at constructing the discrete version of (\ref{Palatini0}) (somewhat more detailed version of the present report is our paper \cite{Kha}).
Let us consider possible discrete version of the Christoffel connection. In the discrete framework, $\Gamma^\lambda_{\mu \nu} \d x^\nu$ corresponds to a finite transformation matrix $\M^\lambda_{\sigma^3 \mu}$ for the transport across some 3-simplex $\sigma^3$. The condition $\Gamma^\lambda_{\mu \nu} = \Gamma^\lambda_{\nu \mu}$ looks unnatural on the discrete level for it requires comparing the matrices $\M^\lambda_{\sigma^3 \mu}$ at the different $\sigma^3$s, that is, is not local. Fortunately, this condition is not required for the GR action being reproduced via equations of motion in the Palatini formalism. Without it, we get for $\Gamma^\lambda_{\mu \nu}$ the unique metric-compatible part $\Gamma^\lambda_{\mu \nu} ( \{ g_{\lambda \mu} \} )$ plus some part $\Gamma_\nu \delta^\lambda_\mu $ with torsion which, however, does not contribute to the action. Thus, we consider the general three-index variable $\Gamma^\lambda_{\mu \nu}$ as the continuum counterpart; the discrete counterpart will be the general nondegenerate transformation matrices $4 \times 4$ on the 3-simplices $\sigma^3$, $\M_{\sigma^3} \in $ GL(4, R).

Just as the Palatini action can be related to the Cartan-Weyl action (that is, the orthogonal connection representation) on the continuum level, we consider possible obtaining some discrete Palatini form of the Regge action by modifying some discrete Cartan-Weyl form of it. We have proposed the latter orthogonal connection representation in the local Euclidean frame formalism in our paper \cite{our}. In this formulation, the triangle $\sigma^2$ is described by any pair of its edge vectors $l^a_{\sigma^1_1}$, $l^a_{\sigma^1_2}$, and the area is $A(\sigma^2 )$ $=$ \\ $\sqrt{ l^2_{\sigma^1_1} l^2_{\sigma^1_2} - (l_{\sigma^1_1} l_{\sigma^1_2})^2} / 2 $. The connection SO(4) matrices $\Omega_{\sigma^3}$ on the 3-simplices $\sigma^3$ are the additional independent variables. The product of these for the set of $\sigma^3$s meeting at $\sigma^2$ ordered along a closed path encircling $\sigma^2$ and passing through each of these (and only these) $\sigma^3$s is the curvature matrix $R_{\sigma^2} (\Omega ) = \prod_{ \{ \sigma^3 : ~ \sigma^3\supset\sigma^2 \} }{\Omega^{\epsilon (\sigma^2, \sigma^3)}_{\sigma^3}}$ on the triangle $\sigma^2$. Here, $\epsilon (\sigma^2, \sigma^3) = \pm 1$ is some sign function. If this path begins and ends in a 4-simplex $\sigma^4$ where $l^a_{\sigma^1_1}$ and $l^a_{\sigma^1_2}$ are defined, the corresponding action takes the form
\begin{equation}                                                            
S^{\rm discr}_{\rm SO(4)} = 2 \sum_{\sigma^2} A(\sigma^2 ) \arcsin \left [ \frac{ R^{ab}_{\sigma^2} ( \Omega ) l^c_{\sigma^1_1} l^d_{\sigma^1_2} }{4 A(\sigma^2 )} \epsilon_{abcd} \right ].
\end{equation}

Now we should transform this expression to the general nondegenerate constant metric $g_{\lambda \mu}$ in each 4-simplex and generalize it from the orthogonal to general nondegenerate connection matrices. To this end, we substitute the edge vectors $l^a$ by the corresponding world coordinate differences $\Delta x^\lambda$, the curvature matrix $R^{ab}_{\sigma^2} ( \Omega )$ by $\R^\lambda_{\sigma^2 \mu} ( \M )$ built of $\M_{\sigma^3}$s just as $R_{\sigma^2} ( \Omega )$ is built of $\Omega_{\sigma^3}$s, and $\epsilon_{abcd}$ by $\epsilon_{\lambda \mu \nu \rho} \sqrt{g}$. As a result, the action turns into
\begin{equation}\label{GL4R-naive}                                          
2 \sum_{\sigma^2} A(\sigma^2 ) \arcsin \left [ \frac{ \R^\lambda_{\sigma^2 \tau} ( \M ) g^{\tau \mu} \Delta x^\nu_{\sigma^1_1} \Delta x^\rho_{\sigma^1_2} }{4 A(\sigma^2 )} \epsilon_{\lambda \mu \nu \rho} \sqrt{g} \right ].
\end{equation}

\noindent Here, the dependence on $\M_{\sigma^3}$ for the given $\sigma^3$ is that one through $\R_{\sigma^2}$ of the form
\begin{equation}                                                            
\R_{\sigma^2} = (\Gamma_1 (\sigma^2 , \sigma^3 ) \M_{\sigma^3} \Gamma_2 (\sigma^2 , \sigma^3 ))^{\epsilon (\sigma^2, \sigma^3)}, ~~~ \sigma^2 \subset \sigma^3
\end{equation}

\noindent where $\Gamma_1 (\sigma^2 , \sigma^3 )$, $\Gamma_2 (\sigma^2 , \sigma^3 )$ are some products of the connection matrices on the 3-faces other than the given $\sigma^3$.

Take for example $\epsilon (\sigma^2, \sigma^3) = + 1$. The contribution to the equations of motion obtained by applying $\M^\nu_{\sigma^3 \lambda} \partial / \partial \M^\nu_{\sigma^3 \mu}$ to this action takes the form
\begin{eqnarray}\label{eq-motion-version}                                   
\hspace{-10mm} \hspace{-10mm} \sum_{ \hspace{10mm} \{\sigma^2 : ~ \sigma^2 \subset \sigma^3 \} } \hspace{-10mm} [ \Gamma_2 (\sigma^2 \! , \! \sigma^3 ) \frac{v_{\sigma^2 } \R_{\sigma^2 }}{\cos \alpha (\! \sigma^2 \! )} \Gamma^{-1}_2 (\sigma^2 \! , \! \sigma^3 ) ]^\mu {}_{ \lambda} = 0, & & \nonumber \\ v_{\sigma^2 \lambda \mu} \! = \! \frac{1}{2} \sqrt{g} \epsilon_{\lambda \mu \nu \rho} \Delta x^\nu_{\! \sigma^1_1} \Delta x^\rho_{\! \sigma^1_2}. \hspace{-10mm} & &
\end{eqnarray}

\noindent where
\begin{equation}                                                            
\alpha (\! \sigma^2 \! ) = \arcsin \left [ \frac{ \R^\lambda_{\sigma^2 \tau} ( \M ) g^{\tau \mu} \Delta x^\nu_{\sigma^1_1} \Delta x^\rho_{\sigma^1_2} }{4 A(\sigma^2 )} \epsilon_{\lambda \mu \nu \rho} \sqrt{g} \right ].
\end{equation}

\noindent If the connection is metric compatible and the curvature rotates around its support, we have the following elementary identities,
\begin{equation}\label{ident}                                               
v_{\sigma^2 } \R^{\pm 1}_{\sigma^2 } + \R^{\mp 1}_{\sigma^2 } v_{\sigma^2 } = 2 v_{\sigma^2 } \cos \alpha (\sigma^2 ). \end{equation}

\noindent It is seen that the equations of motion (\ref{eq-motion-version}) are not reduced to the closure condition for the surface of $\sigma^3$ as it would be in the orthogonal connection representation although resemble the sum of the surface bivectors vanishing.

To make the equations of motion be reducible exactly to the closure condition, we can perform a kind of fine tuning of the definition of the defect angle in terms of the curvature matrix. Namely, the form of the equations of motion (\ref{eq-motion-version}) obtained for the naive version of the action (\ref{GL4R-naive}) and the identities for the metric compatible connection and genuine curvature (\ref{ident}) prompt the following substitution for the curvature matrix to be done in the action (\ref{GL4R-naive}),
\begin{equation}                                                            
\R_{\sigma^2 } \Longrightarrow \frac{1}{2} ( \R_{\sigma^2 } - \R^{-1}_{\sigma^2 } ).
\end{equation}

\noindent That is, the correct discrete action takes the form
\begin{eqnarray}\label{Palatini}                                           
S^{\rm discr}_{\rm GL(4,R)} = 2 \sum_{\sigma^2} A(\sigma^2 ) \arcsin
\left \{ \frac{ [\R_{\sigma^2} - \R^{-1}_{\sigma^2} ]^\lambda {}_{ \tau} ( \M ) g^{\tau \mu} \Delta x^\nu_{\sigma^1_1} \Delta x^\rho_{\sigma^1_2} }{8 A(\sigma^2 )} \epsilon_{\lambda \mu \nu \rho} \sqrt{g} \right \}.
\end{eqnarray}

\noindent We have for the equations of motion instead of (\ref{eq-motion-version})
\begin{eqnarray}                                                           
\hspace{-0mm} \sum_{ \hspace{0mm} \{\sigma^2 : ~ \sigma^2 \subset \sigma^3 \} } \hspace{-0mm} \epsilon (\sigma^2, \sigma^3) \left [ \Gamma_2 (\sigma^2 , \sigma^3 )
\frac{v_{\sigma^2 } \R^{\epsilon (\sigma^2, \sigma^3)}_{\sigma^2 } + \R^{- \epsilon (\sigma^2, \sigma^3)}_{\sigma^2 } v_{\sigma^2 } }{\cos \alpha (\sigma^2 )} \Gamma^{-1}_2 (\sigma^2 , \sigma^3 ) \right ] = 0. & &
\end{eqnarray}

\noindent Taking into account the identity (\ref{ident}), these read
\begin{eqnarray}                                                           
\hspace{-5mm} \sum_{ \hspace{5mm} \{\sigma^2 : ~ \sigma^2 \subset \sigma^3 \} } \hspace{-5mm} \epsilon (\sigma^2, \sigma^3) \left [ \Gamma_2 (\sigma^2 , \sigma^3 ) v_{\sigma^2 } \Gamma^{-1}_2 (\sigma^2 , \sigma^3 ) \right ] = 0.
\end{eqnarray}

\noindent This is the closure condition for the (dual) bivectors of $\sigma^3$ (transported with the help of $\Gamma_2$ to the same point) fulfilled identically.

The piecewise affine coordinate frame is fully described by the set of the coordinates of the vertices (zero dimension simplices $\sigma^0$) $x^\lambda_{\sigma^0}$. This defines the length squared of the edge $\sigma^1$ (1-simplex) with the ending vertices $\sigma^0_1$, $\sigma^0_2$ in terms of the metric $g_{\lambda \mu }$ in any of the 4-simplices containing this $\sigma^1$,
\begin{equation}                                                           
l^2_{\sigma^1 } = ( x^\lambda_{\sigma^0_2 } - x^\lambda_{\sigma^0_1 } ) ( x^\mu_{\sigma^0_2 } - x^\mu_{\sigma^0_1 } ) g_{\lambda \mu }.
\end{equation}

\noindent Vise versa, the coordinates of the vertices and edge lengths define via these equations some metric, constant in each 4-simplex. In overall, the full set of the variables consists of the metric variables (edge lengths $l_{\sigma^1}$ and coordinates of the vertices $x^\lambda_{\sigma^0}$) and connection matrix variables $\M_{\sigma^3} \in {\rm GL(4,R)}$.

The discrete counterpart of the diffeomorphisms of the continuum theory are the piecewise linear transformations of the piecewise affine coordinate frame. These are parameterised by the change of coordinates of the vertices. Correspondingly, the contravariant and covariant vectors are transformed to
\begin{equation}                                                           
\tilde{A}^\lambda = N^\lambda_{\sigma^4 \mu} A^\mu, ~~~ \tilde{A}_\lambda = A_\mu (N^{-1}_{\sigma^4})^\mu {}_\lambda
\end{equation}

\noindent in each 4-simplex $\sigma^4$. Here, the transformation matrix $N^\lambda_{\sigma^4 \mu}$ can be defined from the particular case of these when $A^\lambda$ runs over the edge vectors inside the given $\sigma^4$ emitted from some of its vertices $\sigma^0_0$ to its other vertices $\sigma^0_i, i = 1, 2, 3, 4$,
\begin{equation}                                                           
\tilde{x}_{\sigma^0_i}^\lambda - \tilde{x}_{\sigma^0_0}^\lambda = N^\lambda_{\sigma^4 \mu} ( x_{\sigma^0_i}^\mu - x_{\sigma^0_0}^\mu ).
\end{equation}

In the path integral, the largest contribution comes from the points where the exponential varies slowly. At the large areas $A \gg 1$ (in the Plank scale $10^{-33} cm$ units) this occurs at the small argument $x$ of $\arcsin$ in the action so that $\arcsin x \approx x$. Remind that the curvature enters as $\R - \R^{-1}$. The Haar measure is $\D \M = (\det \M )^{-4} \d^{16} \M $, $\M \in {\rm GL(4,R)}$; as $\R$ depends on $\M$ multiplicatively, it reduces partly to $\D \R$. Roughly, we have a matrix analog of some Bessel function, which is transformed into an absolutely convergent integral by the transition to the contour integration in the complex plane,
\begin{eqnarray}                                                           
\int^{\infty}_0 \exp \left [ iA \left ( x - \frac{1}{x} \right ) \right ] \frac{\d x}{x}
\Longrightarrow \int^{\infty}_0 \exp \left [ - A \left ( x + \frac{1}{x} \right ) \right ] \frac{\d x}{x},
\end{eqnarray}

\noindent $A$ stands for a typical scale of area bivector. The contribution of areas exceeding the Plank scale is probably exponentially suppressed.

To conclude, the affine connection form of the discrete gravity can have interesting implications, some interesting features are the following ones. \\
i) Considerably less variables compared to the local Euclidean/Minkowski frame formalism. For example, the spacetime can be decomposed into the 4-cubes, and each 4-cube can be divided into 24 4-simplices. In the orthogonal connection formalism, we need to set the 24 local frames per 4-cube. In the affine connection formalism, we need only the lengths of the 15 edges and 4 coordinates of the vertex per 4-cube. \\
ii) The same group GL(4,R) for both the Euclidean and Minkowski cases. \\
iii) Explicit mechanism of convergence of the functional integral. \\
iv) There is some specific feature of using the affine connection: the freely chosen elements $\M_{\sigma^3} \in {\rm GL(4,R)}$ do not automatically enter the domain of definition of $S^{\rm discr}_{\rm GL(4,R)}$. We need to specially check that the arguments of $\arcsin$'s are not greater than unity in absolute value.

\section*{Acknowledgements}
The present work was supported by the Ministry of Education and Science of the Russian Federation.

\end{document}